\documentclass[pdflatex,sn-mathphys-num]{sn-jnl}% Math and Physical Sciences Numbered Reference Style
\usepackage[english]{babel}
\usepackage{amsthm, amssymb, mathrsfs, amsmath, stmaryrd }
\usepackage{bold-extra}%
\usepackage{graphicx}%
\usepackage{multirow}%
\usepackage[title]{appendix}%
\usepackage{xcolor}%
\usepackage{textcomp}%
\usepackage{manyfoot}%
\usepackage{booktabs}%
\usepackage{algorithm}%
\usepackage{algorithmicx}%
\usepackage{algpseudocode}%
\usepackage{listings}%
\usepackage{relsize}%
\usepackage{xurl}% allow line breaks anywhere in URLs
\usepackage{hyperref}%
\usepackage[capitalize,noabbrev]{cleveref}%
\usepackage{bbm}%
\usepackage{parskip}%
\usepackage{enumerate}%
\usepackage{svg}%
\usepackage{graphicx}%
\usepackage{hyperref}%
\usepackage{xcolor}%
\usepackage{orcidlink}%
\usepackage[multiple]{footmisc}%
\usepackage{setspace}%
\usepackage{bold-extra}%
\usepackage{subcaption}%

\theoremstyle{thmstylethree}%
\newtheorem{theorem}{Theorem}% meant for continuous numbers
\newtheorem{proposition}{Proposition}%
\newtheorem*{proposition*}{Proposition}

\theoremstyle{thmstyletwo}%

\theoremstyle{thmstylethree}%
\newtheorem{definition}{Definition}%

\newcommand{\bb}[1]{\mathbb{#1}}

\newcommand{\dif}{\mathrm{d}}
\newcommand{\scr}[1]{\mathscr{#1}}

\renewcommand{\cal}[1]{\mathcal{#1}}
\newcommand{\inlinetitle}[2] {\noindent\textbf{#1{#2}}}
\newcommand{\argmin}[1]{\underset{#1}{\arg \min} \;}
\newcommand{\argmax}[1]{\underset{#1}{\arg \max} \;}
\newcommand{\defeq}{\stackrel{\mathsmaller{\mathsf{def}}}{=}}
\renewcommand{\epsilon}{\varepsilon}

\newcommand{\footurl}[1]{\footnote{\url{#1}}}
\renewcommand{\emptyset}{\varnothing}

\makeatletter
\newcommand{\xnrightarrow}[2][]{%
  \mathrel{%
    \vphantom{\xrightarrow[#1]{#2}}%
    \ooalign{\hidewidth\neg@arrow\hidewidth\cr$\m@th\xrightarrow[#1]{#2}$\cr}%
  }%
}
\newcommand{\neg@arrow}{%
  $\m@th\vcenter{\hbox{%
    \rotatebox[origin=c]{-45}{\scalebox{1.5}[1]{$\m@th\scriptscriptstyle|$}}%
  }}$
}
\makeatother

\definecolor{accent}{HTML}{657ed4}
\hypersetup{
  colorlinks=true,
  linkcolor=accent,
  urlcolor={accent},
  filecolor={accent},
}

\usepackage[normalem]{ulem}
\usepackage{color}

\newcounter{marginNoteCounter}

\begin{document}

\title[A Unifying Framework for Global Optimization: From Theory to Formalization]{A Unifying Framework for Global Optimization: From Theory to Formalization}


\author*[1]{\fnm{Gaëtan} \sur{Serré} \orcidlink{0009-0006-3503-2581}}\email{gaetan.serre@ens-paris-saclay.fr}
\author[1]{\fnm{Argyris} \sur{Kalogeratos} \orcidlink{0000-0003-1436-3593}}\email{argyris.kalogeratos@ens-paris-saclay.fr }
\author[1]{\fnm{Nicolas} \sur{Vayatis} \orcidlink{0000-0003-4308-4681}}\email{nicolas.vayatis@ens-paris-saclay.fr }
\affil[1]{\orgdiv{Centre Borelli}, \orgname{École Normale Supérieure Paris-Saclay}, \orgaddress{ \city{Gif-Sur-Yvette}, \postcode{91190}, \country{France}}}


\abstract{We introduce an abstract measure‑theoretic framework that serves as a tool to rigorously study stochastic iterative global optimization algorithms as a unified class. The framework is formulated in terms of probability kernels, which, via the Ionescu–Tulcea theorem, induce probability measures on the space of sequences of algorithm iterations, endowed with two intuitive properties. This framework answers the need for a general, implementation‑independent formalism in the analysis of such algorithms, providing a starting point for formalizing global optimization results in proof-assistants. To illustrate the relevance of our tool, we show that common algorithms fit naturally in the framework, and we also use it to give a rigorous proof of a general consistency theorem for stochastic iterative global optimization algorithms (Proposition~3 of \cite{Malherbe2017}). This proof and the entire framework are formalized in the \texttt{Lean} proof assistant. This formalization both ensures the correctness of the definitions and proofs, and provides a basis for future machine-assisted formalizations in the field.}




\maketitle

\section{Introduction}
Inspired by the broad adoption of optimization techniques across fast‑growing areas such as machine learning, alongside the concurrent rise of proof assistants like Lean \cite{lean} and its Mathlib library \cite{mathlib}, this work tackles a core gap at the intersection of algorithmic analysis and formal verification. Although many stochastic iterative global optimization algorithms have been developed and studied individually \cite{Pinnau2017,Malherbe2017, Carrillo2021, Serre2025}, their systematic formalization in proof assistants remains scarce. While a general kernel‑based perspective has appeared in the literature (e.g., \cite{Gomez2019}), existing approaches lack the development required for mechanized verification; specifically, they do not build probability laws over sequence spaces nor provide a proof‑assistant formalization, both indispensable for machine‑assisted verification. This fragmentation impedes the formalization of general results across families of algorithms, forcing each to be treated in isolation.

Our contribution resolves this obstacle by proposing an abstract, measure‑theoretic framework that unifies the analysis of all stochastic iterative global optimization algorithms within a single kernel‑based formalism. This unified approach enables formal assistants to verify general theorems that hold for entire classes of algorithms rather than single instances, substantially reducing redundant proof work. Crucially, once results are formalized within our framework in a system like \texttt{Lean}, they become reusable components for subsequent formal developments thanks to the inherent modularity of proof assistants. As a result, definitions and properties established within this framework are automatically available for reuse in future formalizations. By providing a rigorous, machine-verifiable foundation for optimization algorithms, this work harnesses the inherent modularity of proof assistants, transforming algorithmic verification from an ad hoc, labor-intensive undertaking into a cumulative practice where each formal contribution strengthens the shared knowledge base.

%\inlinetitle{Global optimization}{.}~%
\textbf{Global optimization} is the problem of finding the global optimizer (maximizer or minimizer) of a real-valued function $f$ over a search space $\Omega$, usually a subset of $\bb{R}^d$. Global optimization problems can be classified into two main categories depending on the countability of $\Omega$: if $\Omega$ is finite or countably infinite, the problem is called {\it discrete global optimization}; if $\Omega$ is uncountable, the problem is called {\it continuous global optimization}. The strategies used in these two categories are very different, and in this paper we focus on continuous global optimization. In this setting, the function $f$ is usually assumed to be continuous, and the search space $\Omega$ is often assumed to be compact, which ensures that $f$ has a global maximum and minimum within $\Omega$. The problem of continuous global optimization can be formally expressed as follows:
$$
  x^\star \in \argmax{x \in \Omega} f(x).
$$
The above is expressed using the $\arg\max$ operator, though it is equivalent to using its complementary $\arg\,\min$:
%$$
  $\argmax{x \in \Omega} f(x) = \argmin{x \in \Omega} (-f(x))$.
%$$

The problem of global optimization is equivalent to that of local optimization only when $f$ is a convex function, which explains why global optimization is often called {\it non-convex optimization}. The strategies employed to solve global optimization problems are different from those used to solve local optimization problems, and the theoretical analysis of global optimization algorithms can be more challenging, as the class of functions considered is much larger and the algorithms often explore stochastically the search space \cite{Hansen1996, Pinnau2017, Frazier2018, Contal2013, Serre2025}.

The applications of global optimization are vast, ranging from machine learning and data science to operations research and engineering design \cite{Pinter1996, lee2017finding, Awasthi2024, Houssein2024, el2024taxonomy, Franceschi2024}. Although the study of global optimization algorithms dates decades back \cite{Kirkpatrick1983,Price1983}, it remains an active research area with many recent contributions, notably on SDE-driven methods \cite{Pinnau2017, Bungert2024, riedl2024leveraging, fornasier2024consensus, gerber2025mean}.

The strategies developed to tackle global optimization problems are highly diverse. To cite a few examples, we can mention \textsc{simulated annealing} \cite{Kirkpatrick1983}, which is inspired by the annealing process in metallurgy; \textit{genetic algorithms} \cite{Holland1984}, which mimic the process of natural selection; \textit{particle swarm optimization} \cite{Kennedy1995}, which is based on the social behavior of birds and fish; \textsc{covariance matrix adaptation evolution strategy} \cite{Hansen1996}, which adapts the covariance matrix of a multivariate normal distribution to sample new candidate solutions. More recent algorithms include \textit{bayesian optimization} \cite{Jones2001, Shahriari2016}, which uses probabilistic models to guide the search for the optimum; \textsc{adalipo} \cite{Malherbe2017}, which combines Lipschitz optimization with random exploration; \textsc{consensus-based optimization} \cite{Pinnau2017} or \textsc{stein boltzmann sampling} \cite{Serre2025}, which uses a system of interacting particles to explore the search space.

%\inlinetitle{Proof assistants}{.}~%
\textbf{Proof assistants} are software for formalizing and mechanically checking mathematical proofs. Proof assistants are built on extensions of the calculus of constructions \cite{Coquand1988}, where propositions are represented as types and proofs as terms inhabiting those types. Thus, proving a statement in a proof assistant amounts to constructing a term of the corresponding type. Multiple proof assistants exist, such as {\tt Rocq} \cite{rocq}, {\tt Isabelle/HOL} \cite{isabelle}, and {\tt Lean} \cite{lean}. The use of proof assistants in foundational mathematics has been growing in recent years, with notable successes including the formalization of the four-color theorem \cite{Gonthier2008} and the Feit-Thompson theorem \cite{Gonthier2013} in {\tt Rocq}; the formalization of the Kepler conjecture \cite{Hales2017} in {\tt Isabelle/HOL}; the formalization of the sphere eversion theorem \cite{Van2023} and the independence of the continuum hypothesis \cite{Han2020} in {\tt Lean}; and more \cite{Wieser2022, bobbin2024formalizing, del2025formalizing}. Recently, {\tt Lean} has gained popularity due to its user-friendly interface and powerful libraries, such as {\tt Mathlib} \cite{mathlib}, which provides a comprehensive collection of undergraduate and graduate-level mathematics, including real analysis, measure theory, and topology. Renowned mathematicians such as Terence Tao have also embraced {\tt Lean} for formalizing their research, including Tao's work on the Polynomial Freiman-Ruzsa conjecture \cite{Gowers2025} and his analysis textbook \cite{Tao2022}. Tao has documented his experience with formalization in a series of blog posts\footurl{https://terrytao.wordpress.com/2023/11/18/formalizing-the-proof-of-pfr-in-lean4-using-blueprint-a-short-tour/}\textsuperscript{,}\footurl{https://terrytao.wordpress.com/2025/05/31/a-lean-companion-to-analysis-i/}. In an online article\footurl{https://terrytao.wordpress.com/wp-content/uploads/2024/03/machine-assisted-proof-notices.pdf}, Terence Tao advocates a collaborative role for proof assistants and machine‑learning algorithms: by grounding ML outputs in formal verification, proof assistants could curb model hallucinations, while ML could suggest promising proof directions and accelerate the formalization process.

Although proof assistants have proven effective in foundational mathematics, their adoption in applied areas such as optimization remains limited. This stems from the complexity of formalizing algorithms and their analyses, the steep learning curve of proof assistants, and the limited overlap between formal-methods and applied-mathematics communities. Paradoxically, applied fields stand to gain the most from mechanized verification. In applied mathematics, where empirical results and algorithmic performance take precedence, proofs are secondary artifacts that rarely undergo the detailed peer review that is standard in pure mathematics. Undetected errors in such proofs can compromise entire lines of research. Formal verification thus addresses a critical gap, offering systematic validation where traditional review processes fall short.

\inlinetitle{Contributions}{.}~%
To this end, we set out to formally verify a non-trivial result from the global optimization literature using \texttt{Lean}. After surveying the field, we identified Proposition~3 from \cite{Malherbe2017} as a suitable candidate. This proposition establishes, for \textbf{any} stochastic iterative global optimization algorithm, an equivalence between the consistency of a stochastic global optimization algorithm and the property of sampling the entire search space.

\begin{proposition}[Proposition~3 from \cite{Malherbe2017}]\label{prop:consistency}
  Let $\scr{A}$ be a stochastic iterative global optimization algorithm. The sequence $(X_0, \dots, X_n)$ denotes a sequence of $n + 1$ points sampled by $\scr{A}$ over a function $f$. The following statements are equivalent:
  \begin{enumerate}[i.]
    \item For any Lipschitz continuous function $f$ defined on a compact measurable metric space $\Omega$,
      $$
        \sup_{x \in \Omega} \min_{i = 0 \dots n} \textnormal{dist}(X_i, x) \xrightarrow{p} 0.
      $$
    \item For any Lipschitz continuous function $f$ defined on a compact measurable metric space $\Omega$,
      $$
        \max_{i = 0 \dots n} f(X_i) \xrightarrow{p} \max_{x \in \Omega} f(x).
      $$
  \end{enumerate}
\end{proposition}

Such general results are rare, and the proof is sufficiently intricate to make it an ideal test case for formal verification. However, the formalization effort revealed a fundamental obstacle: the absence of a rigorous, unifying definition of stochastic iterative global optimization algorithms. Given the diversity of approaches in the field, no general framework existed to support a systematic study, making it difficult to even state the proposition precisely. Some related works have started to explore this direction (e.g. \cite{Powell2019, Casgrain2019, Chou2020}), but they are either limited to specific algorithm classes or lack the formalism needed for implementation‑independent reasoning.

To overcome this challenge, we develop an abstract mathematical framework that serves as a tool to rigorously study any stochastic iterative global optimization algorithm, in a formal way. This framework is based on probability kernels, which are mathematical objects describing the probabilistic transitions between states in a stochastic process. The idea of using probability kernels has the advantage that we can capture the internal mechanisms of any stochastic iterative global optimization algorithm without explicitly modeling it. Our framework allows the construction of probability measures on the spaces of finite and infinite sequences of algorithm iterations, via the Ionescu–Tulcea theorem \cite{Tulcea1949}. We prove two intuitive theorems about these measures that establish key properties useful for rigorous analysis of global optimization algorithms; they also act as sanity checks for the soundness of our framework and serve as exemplars of reasoning within it. Finally, we demonstrate how our framework enables a rigorous proof of Proposition~3 from \cite{Malherbe2017}, which we fully formalize in \texttt{Lean}, providing machine-assisted verification of both the framework and the proposition's proof. We emphasize that the general framework introduced in \cite{Gomez2019} is closely related in proposing a kernel‑based formalism, but it does not construct probability laws on sequence spaces nor provide a formalization or concrete use case, which are key distinctions of our contribution.

%\section*{Notations}
\inlinetitle{Notations}{.}~%
Let $B(x, \epsilon)$ be the open ball with center $x$ and radius $\epsilon > 0$ for some metric space. Let $\cal{P}(\Omega)$ be the set of probability measures over a measurable space $(\Omega, \scr{F})$. For a metric space specified by the context, we denote by $\text{dist}(\cdot,\cdot)$ the distance function on this space. Depending on the context, the symbol ``$\bigotimes$" denotes the product of $\sigma$-algebras, the product between probability kernels, or the product between measures and probability kernels. The former operation is defined as the $\sigma$-algebra generated by the Cartesian product of measurable sets from each $\sigma$-algebra. The second one is defined by:
$$
(\kappa_1 \otimes \kappa_2) (x, A) = \int_y \kappa_2 \left( (x, y), \{z \; | \; (y, z) \in A\} \right) \dif \kappa_1(x, y),
$$
where $\kappa_1$ is a probability kernel from a measurable space ($E_1$) to another ($E_2$), $\kappa_2$ is a probability kernel from the product of these two measurable spaces ($E_1 \otimes E_2$) to another ($E_3$), and $A$ is a measurable set in the product measurable space $E_2 \otimes E_3$. The last operation, the product measures and probability kernels, is defined by:
$$
(\mu \otimes \kappa) (A) = \int_x \kappa \left(x, \{y \; | \; (x, y) \in A\} \right) \dif \mu(x),
$$
where $\mu$ is a measure on a specific measurable space, $\kappa$ is a probability kernel from this measurable space to another one, and $A$ is a measurable set in the product measurable space. For more details on these operations, we refer the reader to \cite{Kallenberg2021}. Finally, we denote by $(X_i)_{i \in I}$ a sequence of points sampled by a stochastic iterative global optimization algorithm applied on a function (clear from the context), for some countable set $I$ (finite or infinite).

\section{A unifying framework}\label{sec:framework}
Given a continuous function $f : \Omega \to \beta$, any stochastic iterative global optimization algorithm $\scr{A}$ operates the same way. At first, it samples a point $X_0 \in \Omega$. This point is chosen according to the internal structure of the algorithm, potentially at random. Since $f$ has not been evaluated yet, the distribution of $X_0$ is the same for any function. Then, the algorithm iteratively samples points $X_n$ in $\Omega$ for $n > 0$. Apart from the structure of $\scr{A}$, the distribution of $X_n$ only depends on the previous iterations and the evaluations of the function $f$ at these points. Thus, $\scr{A}$ can be reduced to two elements in which the internal structure is encapsulated.

\begin{definition}[Stochastic iterative global optimization algorithm]\label{def:algo}
  Given a measurable search space $\Omega$ and an evaluation space $\beta$ (usually $\bb{R}$) associated with a measurable function $f : \Omega \to \beta$ to optimize, a stochastic iterative global optimization algorithm $\scr{A}$ is defined by:
  \begin{enumerate}[i.]
    \item an initial probability measure $\nu \in \cal{P}(\Omega)$ that describes the distribution of the first point $X_0$ sampled by the algorithm, and
    \item \onehalfspacing a sequence $( \kappa_n : \Omega^{n + 1} \times \beta^{n + 1} \rightsquigarrow \Omega )_{n \in \bb{N}}$, such that, for any $n \in \bb{N}$, $\kappa_n$ is a probability kernel on the product space $\Omega^{n + 1} \times \beta^{n + 1}$ that describes the distribution of the next point $X_{n+1}$ given the previous points $X_0, \dots, X_n$ and their evaluations $f(X_0), \dots, f(X_n)$. Given $f$, this corresponds to the conditional probability $\bb{P}(X_{n + 1} | X_0, \dots, X_n)$.
  \end{enumerate}
\end{definition}

%Note that this definition may not admit further simplification, since its constituents are the fundamental building blocks of any stochastic iterative process.

To assess the validity of this definition, we encompass several well-known algorithms in the literature into this framework.

\inlinetitle{\textsc{pure random search}}{}~%
%\textsc{pure random search} 
is one of the simplest stochastic iterative global optimization algorithm, which samples points uniformly at random in the search space $\Omega$. Expressing this algorithm in the frame of \cref{def:algo} is possible by an initial measure $\nu$ that is the uniform measure on $\Omega$, and kernels $\kappa_n$ that are the same for all $n$: they are the constant kernels that return the uniform measure on $\Omega$ for any previous points and evaluations.

\inlinetitle{\textsc{adalipo}}{}~%
%\textsc{adalipo} 
\cite{Malherbe2017} is designed to find the global maximum of a real-valued Lipschitz function $f$. It uses an estimation of the Lipschitz constant of $f$ to adaptively sample the search space. Expressing this algorithm in the frame of \cref{def:algo} can be done by setting the initial measure $\nu$ to be the uniform measure on $\Omega$, and the kernels $\kappa_n$ to be defined as follows:
$$
  \kappa_n (e) = \cal{U}\left(\left\{ x \in \Omega \ \middle|\ \max_{i = 0 \dots n} f(X_i) \le \min_{i = 0 \dots n} f(X_i) + L(e) \cdot \text{dist} (x, X_i) \right\}\right),
$$
where $e \defeq ((X_0, ..., X_n), (f(X_0), ..., f(X_n)))$, $L(e)$ is an estimation of the Lipschitz constant of $f$ based on the previous evaluations, and $\cal{U}(\cdot)$ is the uniform measure. This kernel ensures that the next point $X_{n+1}$ is sampled in a region where $f(X_{n+1})$ is above a linear upper bound based on the previous evaluations.

\inlinetitle{\textsc{cma-es}}{}~%
%\textsc{cma-es} 
\cite{Hansen1996} is a well-known continuous global optimization algorithm. At each iteration, it samples $k$ points according to a multivariate normal distribution where the mean and the covariance matrix depend on the previous iterations. Given a continuous function $f : \bb{R}^d \to \bb{R}$, to encompass this algorithm by \cref{def:algo} we construct $f_k : \Omega^k \to \bb{R}^k$ by:
$$
  f_k(x_1, \dots, x_k) = (f(x_1), \dots, f(x_k)).
$$
This new definition lets us view \textsc{cma-es} as a procedure that generates samples in $\bb{R}^{d \times k}$, where $f_k$ is employed to construct the transition kernels.

The initial measure $\nu$ is then the $k$-product of multivariate normal distributions with mean $m$ and covariance matrix $C$, two parameters of the algorithm. The kernels $\kappa_n$ are defined as follows:
$$
  \kappa_n (e) = \prod_{i=1}^{k} \cal{N}(\mu(e), \Sigma(e)),
$$
where $e \defeq ((X_0, ..., X_n), (f_k(X_0), ..., f_k(X_n)))$ and where $\mu(e) \in \bb{R}^d$ and $\Sigma(e) \in \bb{R}^{d \times d}$ are functions of $e$ that computes a mean vector and a covariance matrix of a multivariate normal distribution in $\bb{R}^d$. These functions use all the previous samples and evaluations; their definitions are intrinsic to \textsc{cma-es} and should not be interpreted as the empirical mean or covariance of $e$.

\inlinetitle{SDE-based algorithms}{.}~Easy to see that any global optimization algorithm that is based on a system of stochastic differential equations (e.g. \cite{Pinnau2017, Xu2018, Dieuleveut2020, Serre2025}), is trivially encompassed by the presented \cref{def:algo}: the initial measure $\nu$ is the initial distribution of the system and, for any $n \in \bb{N}$, the kernel $\kappa_n$ is the transition kernel coming from the Markov process associated to the system (see \cite{LeGall2016, Chaintron2022}). Moreover, if the step size of the numerical scheme used to solve the system is fixed (such as in the Unadjusted Langevin algorithm \cite{Durmus2017}), then the kernels $\kappa_n$ are the same for any $n$.

\subsection{Link with stochastic process}
The foundational idea upon which this definition builds is that an iterative stochastic global optimization algorithm can be seen as a stochastic process. Indeed, provided the search space $\Omega$ is measurable, with $\scr{F}$ being the associated $\sigma$-algebra, one could construct the probability measure $\bb{P}$ on $\Omega^\bb{N}$ that describes the law of infinite sequences of points sampled by the algorithm. The iterations of the algorithm is then a collection of random variables, which can be written as
$$
  \left\{ X_n : \Omega^\bb{N} \to \Omega \; | \; n \in \bb{N} \right\},
$$
and can be seen as a stochastic process with respect to the probability space $\left( \Omega^\bb{N}, \scr{F}^{\otimes \bb{N}}, \bb{P} \right)$. However, constructing $\bb{P}$ (a measure on infinite sequences of $\Omega$) solely from the internal structure of the algorithm (which only provides the distribution of the next point sampled by the algorithm given the previous points and their evaluations) is not straightforward. One way to proceed is to invoke the Ionescu–Tulcea theorem \cite{Tulcea1949}, which operates on a sequence of probability kernels, exactly the data provided by \cref{def:algo}.

The construction of the measure on infinite sequences, denoted $\mu_\infty^{(f)}$ (where $f$ is the function being optimized), is detailed in \cref{sec:measure_construction}, where we show that it arises from a family of finite‑dimensional laws 
%$\left( \mu_n^{(f)} \right)_{n \in \bb{N}}$ 
$\big( \mu_n^{(f)} \big)_{n \in \bb{N}}$, each $\mu_n^{(f)}$ being a probability measure on the space of finite sequences of length $n+1$ generated by the algorithm. These measures are built from the initial distribution $\nu$ and the kernels $(\kappa_n)_{n \in \bb{N}}$ and satisfy two intuitive properties useful for rigorous analysis of global optimization algorithms.

\subsection{Properties of the measures}
In this section, we present two intuitive properties of the measures %$\left( \mu_n^{(f)} \right)_{n \in \bb{N}}$ 
$\big( \mu_n^{(f)} \big)_{n \in \bb{N}}$
 that arise from our framework. They are expected from the way stochastic iterative global optimization algorithms operate, yet their formal proofs require careful measure-theoretic reasoning. They act as sanity checks for the soundness of our framework, confirming that it aligns with intuitive expectations about algorithm behavior, and they also serve as exemplars of reasoning within it. The properties are needed in the formal proof of \cref{prop:consistency}, which is detailed in \cref{sec:sketch_prop}.

\subsubsection{Equality of restricted measures}
\begin{figure}[tb]
\centering
\includegraphics[width=\textwidth]{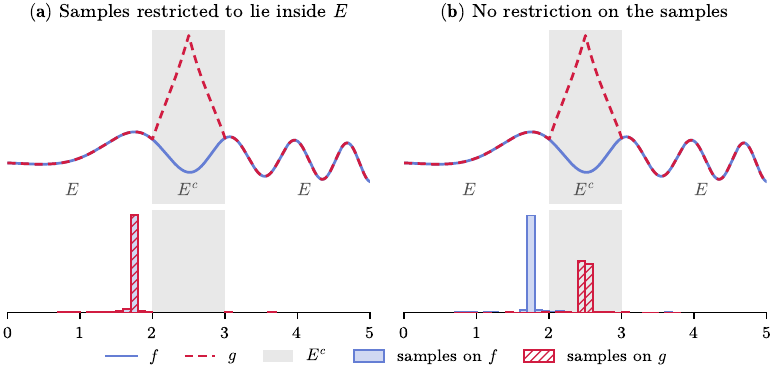}
\caption{Distribution of samples generated by \textsc{cma-es}, when maximizing two different functions $f$ and $g$ that are equal inside the set $E$ (outside the shaded area $E^c$). The histograms of the samples are shown at the bottom. In \textbf{(a)}, we force all samples to lie inside $E$, and we observe that the distributions of samples for both functions are identical. In \textbf{(b)}, we allow samples to lie anywhere in the search space, and we observe that the distributions of samples for both functions differ. This illustrates \cref{thm:eq_restrict}.}
\label{fig:eq_restricted}
\end{figure}

Given two functions $f, g : \Omega \to \bb{R}$ that are equal on a set $E \subseteq \Omega$, a stochastic iterative global optimization algorithm $\scr{A}$ will behave identically if restricted to $E$. More precisely, the measures describing the sequences of points sampled by the algorithm for both functions coincide when we only consider sequences that lie entirely within $E$. This property is stated in the following theorem (see \cref{fig:eq_restricted} for an illustration), and its proof is provided in \cref{sec:proof_eq_restrict}.

\begin{theorem}[Equality of restricted measures]\label{thm:eq_restrict}
  Let $\scr{A}$ be a stochastic iterative global optimization algorithm as in \cref{def:algo}, $E \subseteq \Omega$ a measurable set, $n \in \bb{N}$, and $f, g : \Omega \to \bb{R}$ two measurable functions such that $f(x) = g(x)$ for any $x \in E$. The following holds:
  $$
    \mu_n^{(f)}|_{E^{n + 1}} = \mu_n^{(g)}|_{E^{n + 1}}.
  $$
\end{theorem}

\subsubsection{Monotonicity by truncation}
\begin{figure}[t]
  \centering
  \includegraphics[width=0.7\textwidth]{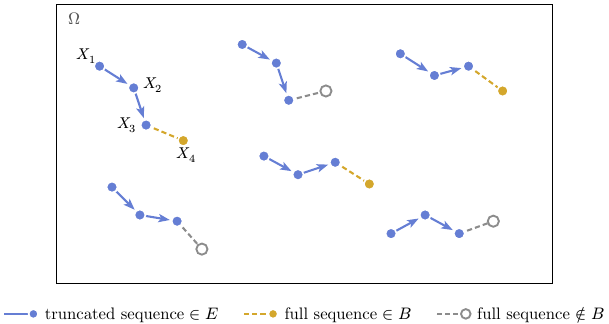}
  \caption{Symbolic illustration of sequences of length $4$, representing algorithmic iterations starting from different initial conditions. For each sequence lying in $B$ (last point connected by a yellow dashed line), its truncated sequence belongs to $E$ (blue points), but the converse does not hold. As a result, there is at least as many sequences of length $3$ in $E$ as there are sequences of length $4$ in $B$. This illustrates \cref{thm:fin_measure_mono}.}
  \label{fig:monotone}
\end{figure}

Another intuitive property is that, if an event defined on the first $m$ iterations implies an event defined on the first $n$ iterations (with $m \le n$), then the probability of the former cannot exceed that of the latter. This intuitive monotonicity is stated in the following theorem (see \cref{fig:monotone} for an illustration).

\begin{theorem}[Monotonicity by sequence truncation]\label{thm:fin_measure_mono}
  Let $\scr{A}$ be a stochastic iterative global optimization algorithm as in \cref{def:algo}, $f : \Omega \to \bb{R}$ a measurable function, $n, m \in \bb{N}$ such that $n \le m$, a measurable set $E \subseteq \Omega^{n + 1}$ and a measurable set $B \subseteq \Omega^{m + 1}$. The following holds:
  $$
    B \subseteq \left\{ (X_i)_{0 \le i \le m} \; \middle| \; (X_i)_{0 \le i \le n} \in E \right\} \implies \mu_m^{(f)} (B) \le \mu_n^{(f)} (E).
  $$
\end{theorem}

The proof is provided in \cref{sec:proof_mono}.

\section{Lean formalization}
The entire framework presented in this paper (including \cref{def:algo,thm:fin_measure_mono,thm:eq_restrict}) has been fully formalized in the \texttt{Lean} proof assistant \cite{lean}. To provide intuition for how to use the framework, we also implement two algorithms as instances of our abstract definition. The framework formalization and these illustrative instances are available in the \texttt{LeanGO} repository\footurl{https://github.com/gaetanserre/LeanGO} with documentation\footurl{https://gaetanserre.fr/LeanGO/}. Building on LeanGO, the proof of \cref{prop:consistency} is formalized in the \texttt{LipoCons} repository\footurl{https://github.com/gaetanserre/LipoCons}, with documentation\footurl{https://gaetanserre.fr/LipoCons/}.

\subsection{Overview of Lean}
\texttt{Lean} \cite{lean} is an open-source proof assistant based on dependent type theory. It is a programming language with a strong type system that allows users to define mathematical objects, state theorems, and construct proofs in a formal language that can be checked by a computer. It is based on the \textit{calculus of inductive constructions} \cite{Coquand1988}, which consider propositions as types and proofs as terms inhabiting those types. Thus, proving a statement in \texttt{Lean} reduces to constructing a term of the corresponding type. To help users constructing such terms, \texttt{Lean} provides a ``tactic mode'', where users can apply high-level commands (called tactics) to manipulate the current proof state. Tactics correspond to common proof techniques in mathematics, such as induction, case analysis, and contradiction. To prove complex mathematical statements, users often combine \texttt{Lean} with the \texttt{Mathlib} library \cite{mathlib}, which provides a comprehensive collection of definitions, theorems, and proofs in various areas of mathematics, including real analysis, measure theory, and topology.

To give a sense of how \texttt{Lean} code looks like, we provide a simple example below. The following \texttt{Lean} code states and proves that the function $x : \bb{R} \mapsto x + 1$ is strictly monotone:

\begin{center}
\includegraphics[]{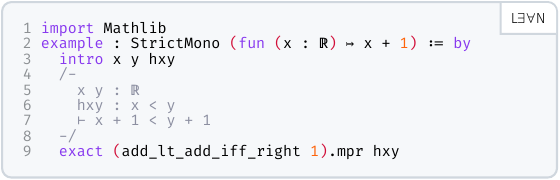}
\end{center}

Let us briefly explain this code. The first line uses the \texttt{import} command to import the definitions of real numbers and of strict monotonicity from \texttt{Mathlib}. The second line states the example, where \texttt{StrictMono} is the proposition that a function is strictly monotone. The third line begins the proof, where \texttt{intro} is a tactic that introduces two arbitrary real numbers $x$ and $y$ such that $x < y$. The proof state is then updated to show that $x + 1 < y + 1$. The fourth line uses a lemma from \texttt{Mathlib} that states that, for any real numbers $x, y, a$, $x + a < y + a \iff x < y$.

\subsection{Formalization of the proposed framework}
The first step of our formalization is to implement the framework presented in \cref{def:algo}. To do so, we need to have \texttt{Lean} definitions of measurable spaces, measures, and kernels. While these definitions are already available in \texttt{Mathlib}, we detail them here for completeness.

\inlinetitle{Measurable space}{.~} A measurable space is a set $X$ equipped with a $\sigma$-algebra $\scr{F}$, which is a collection of subsets of $X$ that includes the empty set, is closed under complements, and is closed under countable unions. In \texttt{Mathlib}, a measurable space is defined as \textit{typeclass}, which is a way to attach additional structure to a type. The \texttt{Mathlib} definition of a measurable space is as follows:

\begin{center}
\includegraphics[]{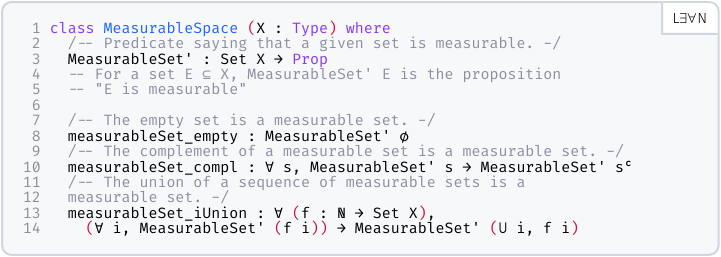}
\end{center}

This typeclass attaches to a type $X$ (which corresponds to a set in classical mathematics) a $\sigma$-algebra by providing a predicate \texttt{MeasurableSet'} that determines whether a subset of $X$ is measurable. It also provides the necessary properties to ensure that \texttt{MeasurableSet'} defines a $\sigma$-algebra.

\inlinetitle{Measure}{.~} A measure on a measurable space $(X, \scr{F})$ is a function $\mu : \scr{F} \to \overline{\bb{R}}_+$ that assigns a non-negative extended real number to each measurable set, satisfying the properties of non-negativity, null empty set, and countable additivity. In \texttt{Mathlib}, a measure is defined as a \textit{structure}, which is a way to group related data together (such as a function and its properties). The \texttt{Mathlib} definition of a measure is rather involved: it is defined as an outer measure (a function from sets to $\overline{\bb{R}}_+$ that sends $\emptyset$ to $0$ and is countably subadditive) that is countably additive on measurable sets. The \texttt{Mathlib} definition of a measure is as follows:

\begin{center}
\includegraphics[]{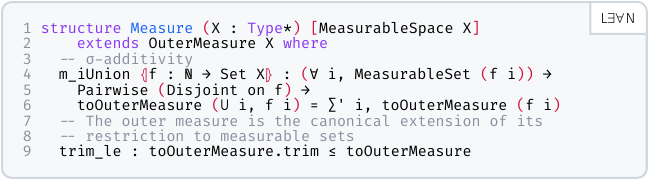}
\end{center}

This structure defines a new type ``$\,$\texttt{Measure} $X$'' for measures on a measurable space $X$. It can be used to create an object $\mu$ that represents a measure on $X$:

\begin{center}
\includegraphics[]{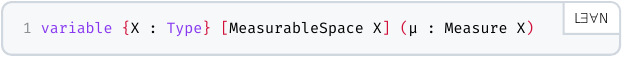}
\end{center}

\inlinetitle{Kernel}{.~} A kernel from a measurable space $(X, \scr{F})$ to another measurable space $(Y, \scr{G})$ is a function $\kappa : X \times \scr{G} \to \overline{\bb{R}}_+$ such that, \textbf{i)} for each fixed $x \in X$, $\kappa(x, \cdot)$ is a measure on $(Y, \scr{G})$, and \textbf{ii)} for each fixed $A \in \scr{G}$, $\kappa(\cdot, A)$ is a measurable function on $(X, \scr{F})$. In \texttt{Mathlib}, a kernel is also defined as a \textit{structure}. The \texttt{Mathlib} definition of a kernel is as follows~\cite{degenne2025markovkernel}:

\begin{center}
\includegraphics[]{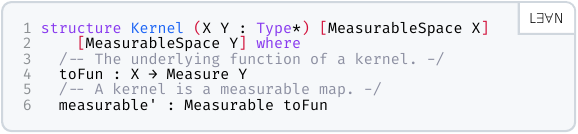}
\end{center}
Note that in this definition, \textbf{ii)} is expressed using the fact that the space of measures on $(Y, \scr{G})$ is itself a measurable space~\cite{Giry1982}.

\inlinetitle{Formalization of \cref{def:algo}}{.~} With these definitions in place, we can now formalize \cref{def:algo} in \texttt{Lean}. A stochastic iterative global optimization algorithm is represented as a \texttt{Lean} structure that encapsulates the initial measure and the sequence of kernels:

\begin{center}
\includegraphics[]{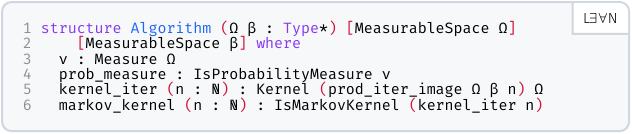}
\end{center}

Here, $\Omega$ is the search space, $\beta$ is the evaluation space (most likely $\bb{R}$), and $\nu$ is the initial measure. The instances \texttt{prop\_measure} and \texttt{markov\_kernel} ensure that $\nu$ is a probability measure and that, for any $n \in \bb{N}$, \texttt{kernel\_iter n} is a probability kernel.

To define the probability measure $\mu_\infty^{(f)}$ of an \texttt{Algorithm}, we used the existing {\tt Lean} formalization of the Ionescu-Tulcea theorem \cite{Marion2025}:
\begin{center}
\includegraphics[]{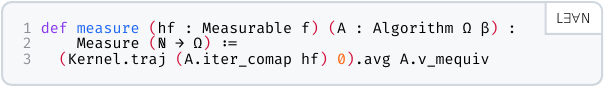}
\end{center}

We do not go into the details of this \texttt{Lean} definition, but it essentially corresponds to the measure constructed using the Ionescu-Tulcea theorem, where ~\href{https://leanprover-community.github.io/mathlib4_docs/Mathlib/Probability/Kernel/IonescuTulcea/Traj.html#ProbabilityTheory.Kernel.traj}{\texttt{Kernel.traj}} is the infinite product of the algorithm's kernels, and where \texttt{iter\_comap} is a technical construction that adapts the kernels to the function $f$ being optimized (see \cref{sec:measure_construction}).

Then, we define the finite-dimensional measures $\left( \mu_n^{(f)} \right)_{n \in \bb{N}}$ by pushing forward $\mu_\infty^{(f)}$ along the function that truncates an infinite sequence to its first $n + 1$ elements:
\begin{center}
\includegraphics[]{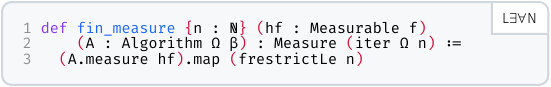}
\end{center}
where \texttt{iter} $\Omega$ \texttt{n} is the \texttt{Lean} type that represents the product space $\Omega^{n + 1}$, and where ~\href{https://leanprover-community.github.io/mathlib4_docs/Mathlib/Order/Restriction.html#Preorder.frestrictLe}{\texttt{frestrictLe n}} is the truncation function.

\inlinetitle{Some examples}{.~} To give a sense of how these definitions look like in practice, we provide the \texttt{Lean} formalization of the \textsc{pure random search} (\textsc{prs}) and the \textsc{lipo}~\cite{Malherbe2017} algorithms. The implementation of these algorithms is available in the \texttt{LeanGO} repository\footurl{https://github.com/gaetanserre/LeanGO}.

\inlinetitle{PRS}{.~} The initial measure $\nu$ of \textsc{prs} is the uniform measure on $\Omega$, and the kernels $\kappa_n$ are the constant kernels that return the uniform measure on $\Omega$ for any previous points and evaluations:
\begin{center}
\includegraphics[]{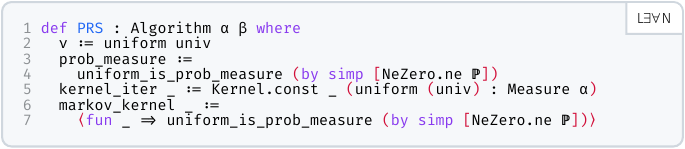}
\end{center}
where $\Omega$ is a measure space with finite and non-zero measure (e.g. a compact subset of $\bb{R}^d$), and where \texttt{uniform} is the function that returns the uniform measure on a given set.

\inlinetitle{LIPO}{.~} The initial measure $\nu$ of \textsc{lipo} is also the uniform measure on $\Omega$, but the kernels $\kappa_n$ are defined as follows:
$$
  \kappa_n (e) = \cal{U}\left(\left\{ x \in \Omega \ \middle|\ \max_{i = 0 \dots n} f(X_i) \le \min_{i = 0 \dots n} f(X_i) + \kappa \cdot \text{dist} (x, X_i) \right\}\right),
$$
where $e \defeq ((X_0, ..., X_n), (f(X_0), ..., f(X_n)))$ and $\kappa$ is the Lipschitz constant of $f$.
The \texttt{Lean} formalization of \textsc{lipo} is more involved as \textsc{prs} because of the definition of the kernels, but it follows the same structure as the formalization of \textsc{prs}:
\begin{center}
\includegraphics[]{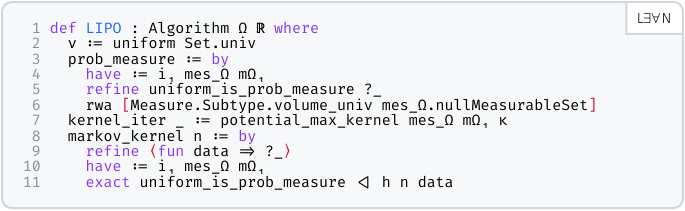}
\end{center}
where $\Omega$ is a subset of $\bb{R}^d$ with finite and non-zero measure, $\kappa$ is a positive real number, and where \texttt{potential\_max\_kernel} is the function that returns the kernel defined by the formula above. We advise the reader to consult the \texttt{Lean} code for the precise definition of \texttt{potential\_max\_kernel}, as it involves several technical constructions to ensure that it is indeed a kernel.

\inlinetitle{Formalization of the properties}{.~}
Finally, we formalized and proved \cref{thm:eq_restrict} and \cref{thm:fin_measure_mono} in {\tt Lean}, both of which were essential for establishing \cref{prop:consistency}.

The \texttt{Lean} statement of \cref{thm:fin_measure_mono} is as follows:
\begin{center}
\includegraphics[]{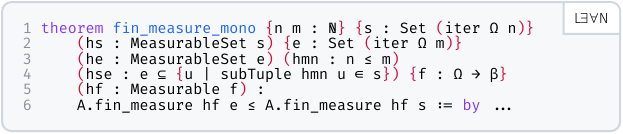}
\end{center}
The \texttt{Lean} proof essentially follows the same reasoning as the proof presented in \cref{sec:proof_mono}.

The \texttt{Lean} statement of \cref{thm:eq_restrict} is as follows:
\begin{center}
\includegraphics[]{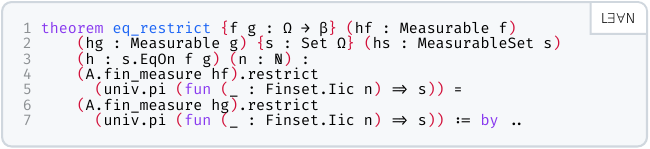}
\end{center}
Analogously, the \texttt{Lean} proof mirrors the proof presented in \cref{sec:proof_eq_restrict}, although the decomposition of the product of kernels is carried out with considerably greater technical precision.

\subsection{Formalization of Proposition 1}
We do not delve into the details of the formalization of the equivalent properties in \cref{prop:consistency}. However, we provide a detailed sketch of the proof in \cref{sec:sketch_prop}, that can be followed alongside the \texttt{Lean} formalization. The \texttt{Lean} statement of \cref{prop:consistency} is as follows:
\begin{center}
\includegraphics[]{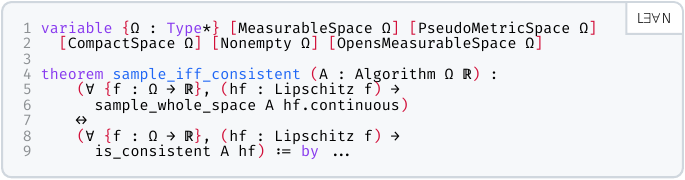}
\end{center}
In this statement, \texttt{sample\_whole\_space} corresponds to property {\bf (i)} of \cref{prop:consistency}, while \texttt{is\_consistent} corresponds to property {\bf (ii)}. The convergence in probability is expressed using \texttt{fin\_measure}, defined in the previous section.

One notable aspect of this formalization is the use of various typeclasses to represent the mathematical structures involved. The search space $\Omega$ is equipped with several typeclasses to ensure it has the necessary properties (measurable space, metric space, compactness, etc.). This approach allows us to work in a general setting without committing to a specific type for $\Omega$.

\section{Formal constructions and proofs}
This section details the construction of the infinite‑sequence measure $\mu_\infty^{(f)}$ introduced in \cref{sec:framework}, gives rigorous proofs of \cref{thm:eq_restrict} and \cref{thm:fin_measure_mono}, and closes with a sketch of the proof of \cref{prop:consistency}.

\subsection{Construction of finite and infinite dimensional laws}\label{sec:measure_construction}
We construct probability measures on finite and infinite sequences sampled by an algorithm $\scr{A}$ as in \cref{def:algo}. To do so, we invoke the Ionescu–Tulcea theorem \cite{Tulcea1949}, which allows us to build such measures from the initial distribution $\nu$ and the sequence of kernels $(\kappa_n)_{n \in \bb{N}}$.

\begin{theorem}[Ionescu-Tulcea theorem \cite{Tulcea1949}]\label{thm:ionescu_tulcea}
  Let $(\Omega_0, \cal{A}_0, P_0)$ be a probability space and $(\Omega_n, \cal{A}_n)_{n \in \bb{N}}$ a sequence of measurable spaces. For each $n \in \bb{N}^*$, let $\kappa_n : \left( \Omega^{n - 1}, \cal{A}^{n - 1} \right) \rightsquigarrow (\Omega_n, \cal{A}_n)$ be a probability kernel, where:
  $$
    \Omega^n \defeq \prod_{k = 0}^n \Omega_k \ \text{ and }\ \cal{A}^n \defeq \bigotimes_{k = 0}^n \cal{A}_k.
  $$
  Then, for each $n \in \bb{N}$, there exists a probability measure $P_n$ defined on $\left( \Omega^n, \cal{A}^n \right)$ by the following formula:
  $$
    P_n \defeq P_0 \otimes \bigotimes_{k = 1}^n \kappa_k,
  $$
  and there exists a uniquely defined probability measure $\bb{P}$ on $\left( \Omega^\bb{N}, \bigotimes\limits_{k = 1}^\infty \cal{A}_k \right)$, so that
  $$
    P_n (A) = \bb{P}\left( A \times \prod_{k = n + 1}^\infty \Omega_k \right).
  $$
  Thus, the measure $\bb{P}$ has conditional probabilities equal to the stochastic kernels $\kappa$.
\end{theorem}

In \cref{def:algo} we are given the initial law of $X_0$ and kernels $\kappa_n$ defined on $\Omega^{n+1}\times\bb{R}^{n+1}$. The Ionescu–Tulcea theorem, however, requires kernels on $\Omega^{n+1}$. To bridge this gap we precompose each $\kappa_n$ along the evaluation map
$$
  e_f : \Omega^{n+1}\to \Omega^{n+1}\times\bb{R}^{n+1},\qquad
  e_f(x_0,\dots,x_n)=\left( (x_0,\dots,x_n),(f(x_0),\dots,f(x_n)) \right),
$$
obtaining pullback kernels on $\Omega^{n+1}$ which are then used in the Ionescu–Tulcea construction.

\begin{definition}[Pullback kernels on finite sequences]
  Given a topological space $\Omega$ endowed with a $\sigma$-algebra $\scr{F}$, a natural number $n$, and a measurable function $f : \Omega \to \bb{R}$. We define $\kappa'^{(f)}_n : \Omega^{n+1} \rightsquigarrow \Omega$ as the pullback of $\kappa_n$ along $e_f$, i.e., for any $x \in \Omega^{n + 1}$,
    $$
      \kappa'^{(f)}_n (x, \cdot) \defeq \kappa_n (e_f(x), \cdot).
    $$
\end{definition}

These new kernels enable the construction of probability measures on finite sequences of points sampled by an algorithm driven by a target function $f$. Indeed, for any $n \in \bb{N}$, a probability measure $\mu_n^{(f)}$ (the notation here emphasizes the dependence on $f$) can be defined on $\left( \Omega^{n + 1}, \bigotimes_{k = 0}^n \scr{F} \right)$ by:
$$
  \mu_n^{(f)} \defeq \nu \otimes \bigotimes_{k = 0}^{n - 1} \kappa'^{(f)}_k.
$$
Moreover, using \cref{thm:ionescu_tulcea}, we can obtain $\mu_\infty^{(f)}$, which is the probability measure on infinite sequences of points sampled by the algorithm over the function $f$ such that:
\begin{equation}\label{eq:finite_measures}
\forall n \in \bb{N}, A \in \bigotimes_{k = 0}^n \scr{F}, \mu_n^{(f)} (A) = \mu_\infty^{(f)} \left( A \times \prod_{k = n + 2}^\infty \Omega \right).
\end{equation}
Therefore, each measure $\mu_n^{(f)}$ is the restriction of $\mu_\infty^{(f)}$ to the first $n + 1$ elements of infinite sequences.

\subsection{Proof of \cref{thm:fin_measure_mono}}\label{sec:proof_mono}
\begin{proof}
  The proof is trivial using \cref{eq:finite_measures}. We have that:
  \begin{align*}
    \mu_m^{(f)}(B) &= \mu_\infty^{(f)} \left( B \times \prod_{k = m + 1}^\infty \Omega \right) \\
    & = \mu_\infty^{(f)} \left( \left\{ (X_i)_{i \in \bb{N}} \; \middle| \; (X_i)_{0 \le i \le m} \in B \right\} \right).
  \end{align*}
  Moreover, for any $X \in \Omega^\bb{N}$ such that $(X_i)_{0 \le i \le m} \in B$, we have by hypothesis that $(X_i)_{0 \le i \le n} \in E$. Thus,
  $$
  \left\{ (X_i)_{i \in \bb{N}} \; \middle| \; (X_i)_{0 \le i \le m} \in B \right\} \subseteq \left\{ (X_i)_{i \in \bb{N}} \; \middle| \; (X_i)_{0 \le i \le n} \in E \right\},
  $$
  and therefore
  \begin{align*}
    \mu_\infty^{(f)} \left( \left\{ (X_i)_{i \in \bb{N}} \; \middle| \; (X_i)_{0 \le i \le m} \in B \right\} \right) &\le \mu_\infty^{(f)} \left( \left\{ (X_i)_{i \in \bb{N}} \; \middle| \; (X_i)_{0 \le i \le n} \in E \right\} \right) \\
    & = \mu_n^{(f)} (E).
  \end{align*}
\end{proof}

\subsection{Proof of \cref{thm:eq_restrict}}\label{sec:proof_eq_restrict}
\begin{proof}~\\
  \inlinetitle{$\mathbf{n = 0}$}{.}~%
  We have that $\mu_0^{(f)} = \nu = \mu_0^{(g)}$, and thus the property holds.

  \inlinetitle{$\mathbf{n \ge 1}$}{.}~%
  It is sufficient to show that the two measures are equal on all measurable rectangles of the form $\prod_{i = 0}^n B_i$ where $(B_i)_{0 \le i \le n}$ are measurable subsets of $\Omega$. Indeed, the $\sigma$-algebra $\bigotimes_{i = 0}^n \scr{F}$ is generated by these rectangles. Moreover, these rectangles form a $\pi$-system. Thus, the Sierpiński–Dynkin's $\pi$-$\lambda$ theorem ensures that two measures are equal on this $\sigma$-algebra if and only if they are equal on all these rectangles.

  Let $(B_i)_{0 \le i \le n}$ be measurable subsets of $\Omega$. The goal reduces to show that
  \begin{align*}
    \mu_n^{(f)}|_{E^{n + 1}} \left( \prod_{i = 0}^n B_i \right) &= \mu_n^{(g)}|_{E^{n + 1}} \left( \prod_{i = 0}^n B_i \right) \\
    \mu_n^{(f)} \left( \prod_{i = 0}^n B_i \cap E \right) &= \mu_n^{(g)} \left( \prod_{i = 0}^n B_i \cap E \right).
  \end{align*}
  For any $i \in \llbracket 0, n \rrbracket$, let $C_i \defeq B_i \cap E$. Using the definition of $\mu_n^{(f)}$, we have that
  \begin{align*}
    \mu_n^{(f)} \left( \prod_{i = 0}^n B_i \cap E \right) =& \left( \nu \otimes \bigotimes_{k = 0}^{n - 1} \kappa'^{(f)}_k \right) \left( \prod_{i = 0}^n C_i \right) \\
    =& \int_x \bigotimes_{k = 0}^{n - 1} \kappa'^{(f)}_k \left(x, \left\{ (y_i)_{1 \le i \le n} \; \middle| \; (x, y) \in \prod_{i = 0}^n C_i \right\} \right) \; \dif \nu(x) \\
    =& \int_{x \in C_0} \bigotimes_{k = 0}^{n - 1} \kappa'^{(f)}_k \left(x, \left\{ (y_i)_{1 \le i \le n} \; \middle| \; (x, y) \in \prod_{i = 0}^n C_i \right\} \right) \; \dif \nu(x) \\
    &+ \int_{x \notin C_0} \bigotimes_{k = 0}^{n - 1} \kappa'^{(f)}_k \left(x, \left\{ (y_i)_{1 \le i \le n} \; \middle| \; (x, y) \in \prod_{i = 0}^n C_i \right\} \right) \; \dif \nu(x) \\
    =& \int_{x \in C_0} \bigotimes_{k = 0}^{n - 1} \kappa'^{(f)}_k \left(x, \left\{ (y_i)_{1 \le i \le n} \; \middle| \; (x, y) \in \prod_{i = 0}^n C_i \right\} \right) \; \dif \nu(x) \\
    &+ \int_{x \notin C_0} \bigotimes_{k = 0}^{n - 1} \kappa'^{(f)}_k \left(x, \emptyset \right) \; \dif \nu(x) \\
    =& \int_{x \in C_0} \bigotimes_{k = 0}^{n - 1} \kappa'^{(f)}_k \left(x, \left\{ (y_i)_{1 \le i \le n} \; \middle| \; (x, y) \in \prod_{i = 0}^n C_i \right\} \right) \; \dif \nu(x) + 0
  \end{align*}
  and similarly,
  $$
  \mu_n^{(g)} \left( \prod_{i = 0}^n B_i \cap E \right) = \int_{x \in C_0} \bigotimes_{k = 0}^{n - 1} \kappa'^{(g)}_k \left(x, \left\{ (y_i)_{1 \le i \le n} \; \middle| \; (x, y) \in \prod_{i = 0}^n C_i \right\} \right) \; \dif \nu(x).
  $$
  Thus, it is sufficient to show that the integrands are equal for any $x \in C_0$.
  Let $A_n \defeq \prod_{i = 1}^n C_i$. For any $x \in C_0$, We have that
  \begin{equation}\label{eq:set_equality}
    \left\{ (y_i)_{1 \le i \le n} \; \middle| \; (x, y) \in \prod_{i = 0}^n C_i \right\} = A_n.
  \end{equation}
  It is sufficient to show that the restriction to $A_n$ of these two kernels products are equal, i.e.
  $$
  \left. \left( \bigotimes_{k = 0}^{n - 1} \kappa'^{(f)}_k (x, \cdot) \right) \right|_{A_n} = \left. \left( \bigotimes_{k = 0}^{n - 1} \kappa'^{(g)}_k (x, \cdot) \right) \right|_{A_n}.
  $$
  Indeed, if this holds, then
  $$
  \bigotimes_{k = 0}^{n - 1} \kappa'^{(f)}_k (x, A_n) = \bigotimes_{k = 0}^{n - 1} \kappa'^{(f)}_k (x, A_n).
  $$
  Combining this with \cref{eq:set_equality}, we obtain the desired result. \\

  We proceed by induction on $n$. For $n = 1$, we have that
  $$
  \kappa'^{(f)}_0 (x, \cdot)|_{A_1} = \kappa_0 (f(x), \cdot)|_{A_1} \text{ and } \kappa'^{(g)}_0 (x, \cdot)|_{A_1} = \kappa_0 (g(x), \cdot)|_{A_1}
  $$
  As $C_0 \subseteq E$ and $f$ and $g$ agree on $E$, we have that $f(x) = g(x)$, and thus
  $$
  \kappa'^{(f)}_0 (x, \cdot)|_{A_1} = \kappa'^{(g)}_0 (x, \cdot)|_{A_1}.
  $$
  Now, let $n \in \bb{N}$ such that the property holds. We want to show that it also holds for $n + 1$. For any measurable set $B \subseteq \Omega^{n + 1}$, we have that
  \begin{align*}
    \left. \left( \bigotimes_{k = 0}^n \kappa'^{(f)}_k (x, B) \right) \right|_{A_{n + 1}}& \\
    = \int_y &\kappa'^{(f)}_n ((x, y), \{z \; | \; (y, z) \in B\})|_{C_{n + 1}} \; \dif \left. \left( \bigotimes_{k = 0}^{n - 1} \kappa'^{(f)}_k (x, y) \right) \right|_{A_n} \\
    = \int_{y \in A_n} &\kappa'^{(f)}_n ((x, y), \{z \; | \; (y, z) \in B\})|_{C_{n + 1}} \; \dif \left. \left( \bigotimes_{k = 0}^{n - 1} \kappa'^{(g)}_k (x, y) \right) \right|_{A_n}.
  \end{align*}
  Moreover, as $x \in C_0$, for any $y \in A_n$, we have that $(x, y) \in E^{n + 1}$, and thus $f((x, y)) = g((x, y))$. Therefore,
  $$
   \kappa'^{(f)}_n ((x, y), \{z \; | \; (y, z) \in B\})|_{C_{n + 1}} = \kappa'^{(g)}_n ((x, y), \{z \; | \; (y, z) \in B\})|_{C_{n + 1}},
  $$
  which concludes the induction and the proof.
\end{proof}

\subsection{Proof sketch of \cref{prop:consistency}}\label{sec:sketch_prop}

In this section, we briefly discuss the strategy of the proof of \cref{prop:consistency} developed in the preprint \cite{Malherbe2017ArXiv} of the original paper \cite{Malherbe2017}, considering that stochastic and iterative global optimization algorithms are defined as in \cref{def:algo}. We do not detail the entire proof, which is already well explained in that paper, but rather highlight the key steps where the framework developed in the present work is essential.

\inlinetitle{Modus ponens}{.}~%
The right implication of \cref{prop:consistency} is quite intuitive. Indeed, if the algorithm samples the search space, then it means that, for any point in this space, the probability for a point sampled by the algorithm to not be arbitrarily close to this point goes to $0$. Therefore, for any $\epsilon > 0$, it suffices to select a $\delta > 0$ such that, for any point $x$ in the search space such that the distance between $x$ and the maximizer of $f$ (noted $x^\star$) is lower than $\delta$, then $f(x^\star) - f(x) \le \epsilon$. This $\delta$ can be obtained using the continuity of $f$. Thus, for any $n \in \bb{N}$, we have the following set inclusion:
$$
  \left\{ (X_i)_{0 \le i \le n} \; \middle| \; f(x^\star) - \max_{i = 0 \dots n} f(X_i) > \epsilon \right\} \subseteq \left\{ (X_i)_{0 \le i \le n} \; \middle| \; \sup_{x \in \Omega} \min_{i = 0 \dots n} \text{dist}(X_i, x) > \delta \right\}.
$$

This allows us to show that, for any $n \in \bb{N}$,
\begin{align*}
  \mu_n^{(f)} &\left( \left\{ (X_i)_{0 \le i \le n} \; \middle| \; f(x^\star) - \max_{i = 0 \dots n} f(X_i) > \epsilon \right\} \right) \le\\
  &\mu_n^{(f)} \left( \left\{ (X_i)_{0 \le i \le n} \; \middle| \; \sup_{x \in \Omega} \min_{i = 0 \dots n} \text{dist}(X_i, x) > \delta \right\} \right),
\end{align*}
by the monotonicity of measures. Then, we can conclude that the left-hand side also goes to $0$ as $n$ goes to infinity using the squeeze theorem, as we know by hypothesis that
$$
  \mu_n^{(f)} \left( \left\{ (X_i)_{0 \le i \le n} \; \middle| \; \sup_{x \in \Omega} \min_{i = 0 \dots n} \text{dist}(X_i, x) > \delta \right\} \right) \xrightarrow[n \to \infty]{} 0.
$$

\inlinetitle{Reverse modus ponens}{.}~%
The left implication is more involved and is proven by contradiction, i.e. we prove that the facts that the algorithm is consistent on any Lipschitz function and that there exists a Lipschitz function $f$ and an $\epsilon_1 > 0$ such that
$$
  \exists \epsilon_2 > 0, \forall N \in \bb{N}, \exists n \ge N, \epsilon_2 < \mu_n^{(f)} \left( \left\{ (X_i)_{0 \le i \le n} \; \middle| \; \sup_{x \in \Omega} \min_{i = 0 \dots n} \text{dist}(X_i, x) > \epsilon_1 \right\} \right)
$$
are incompatible.

First, we use the fact that $\Omega$ is a compact space to construct an $(\epsilon_1/2)$-cover of $\Omega$, i.e. a finite collection of points $\{c_1, \dots, c_{N_1}\}$ in $\Omega$ such that, $\bigcup_{i = 1 \dots N_1} B(c_i, \epsilon_1 / 2) = \Omega$.

Next, we use the contradiction hypothesis, \cref{thm:fin_measure_mono}, and the monotonicity of measures to show that there exists a $c^\star \in \{c_1, \dots, c_{N_1}\}$ such that, for any $n \in \bb{N}$,
\begin{equation}\label{eq:almost_never}
  0 < \frac{\epsilon_2}{2 N_1} \le \mu_n^{(f)} \left( \left\{ (X_i)_{0 \le i \le n} \; \middle| \; \forall i \in \llbracket 0, n \rrbracket, X_i \notin B(c^\star, \epsilon_1 / 2) \right\} \right).
\end{equation}
\begin{proof}
To do so, we again reason by contradiction, assuming that for any $c_i$ in the cover, there exists an $n_i \in \bb{N}$ such that
\begin{equation}\label{eq:le_frac}
\mu_{n_i}^{(f)} \left( \left\{ (X_j)_{0 \le j \le n_i} \; \middle| \; \forall j \in \llbracket 0, n_i \rrbracket, X_j \notin B(c_i, \epsilon_1 / 2) \right\} \right) < \frac{\epsilon_2}{2 N_1}.
\end{equation}
Then, we select $n^\star$, the maximum of all these $n_i$'s. Using \cref{thm:fin_measure_mono}, we show that, for any $c_i$ in the cover,
$$
\mu_{n^\star}^{(f)} \left( \left\{ (X_j)_{0 \le j \le n^\star} \; \middle| \; \forall j \in \llbracket 0, n^\star \rrbracket, X_j \notin B(c_i, \epsilon_1 / 2) \right\} \right) < \frac{\epsilon_2}{2 N_1}.
$$
Indeed, as, for any $c_i$, there exists a $n_i \le n^\star$ such that \cref{eq:le_frac} holds and as the following set inclusion holds:
\begin{align*}
\{ (X_j)_{0 \le j \le n^\star} \; | \; \forall j \in \llbracket 0, n^\star \rrbracket, X_j \notin &B(c_i, \epsilon_1 / 2) \} \subseteq\\
&\{ (X_j)_{0 \le j \le n_i} \; | \; \forall j \in \llbracket 0, n_i \rrbracket, X_j \notin B(c_i, \epsilon_1 / 2) \},
\end{align*}
we can use \cref{thm:fin_measure_mono} to show that, for any $c_i$ in the cover,
\begin{equation}\label{eq:mu_n_star}
\mu_{n^\star}^{(f)} (\{ (X_j)_{0 \le j \le n^\star} \; | \; \forall j \in \llbracket 0, n^\star \rrbracket, X_j\notin B(c_i, \epsilon_1 / 2) \}) < \frac{\epsilon_2}{2 N_1}.
\end{equation}
Finally, showing that
\begin{align*}
\left\{ (X_i)_{0 \le i \le n^\star} \; \middle| \; \sup_{x \in \Omega} \min_{i = 0 \dots n^\star} \text{dist}(X_i, x) > \epsilon_1 \right\} &\subseteq\\
\bigcup_{i = 1 \dots N_1} \{ (X_j)_{0 \le j \le n^\star}& \; | \; \forall j \in \llbracket 0, n^\star \rrbracket, X_j \notin B(c_i, \epsilon_1 / 2) \}
\end{align*}
and using \cref{eq:mu_n_star} %along 
with the $\sigma$-additivity of measures, we obtain the %following 
contradiction:
$$
\mu_{n^\star}^{(f)} \left( \left\{ (X_i)_{0 \le i \le n^\star} \; \middle| \; \sup_{x \in \Omega} \min_{i = 0 \dots n^\star} \text{dist}(X_i, x) > \epsilon_1 \right\} \right) \le \epsilon_2.
$$
\end{proof}
Next, we construct a Lipschitz function $\tilde{f}$ that is indistinguishable from $f$ outside the ball $B(c^\star, \epsilon_1 / 2)$ and such that its maximum is strictly greater than the maximum of $f$ over $\Omega$. This function is constructed as follows:
$$
  \tilde{f}(x) = \begin{cases}
    f(x) + 2 \left( 1 - \frac{\text{dist}(c^\star, x)}{\epsilon_1 / 2} \right) \times \left( \max\limits_{x \in \Omega} f(x) - \min\limits_{x \in \Omega} f(x) + 1 \right) &\text{if } x \in B(c^\star, \epsilon_1 / 2); \\
    f(x) &\text{ otherwise}.
  \end{cases}
$$

\begin{figure}[b]
\centering
\begin{minipage}[b]{0.45\linewidth}
   \centering
   \includegraphics[width=0.9\textwidth]{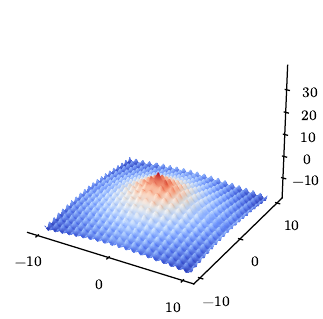} \\
   \vspace*{-0.5em}
   {\small (a) The reverse Ackley function.}
  \end{minipage}
\hfill
\begin{minipage}[b]{0.45\linewidth}
   \centering
   \includegraphics[width=0.9\textwidth]{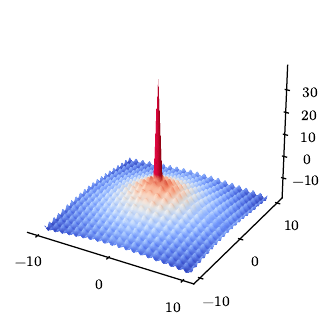} \\
   \vspace*{-0.5em}
   {\small (b) $\tilde{f}$ for the reverse Ackley function.}
\end{minipage}
\vspace*{0.5em}
\caption{Visualization of $\tilde{f}$ applied to the reverse Ackley function on $[-10, 10]^2$ with $c \defeq (0, 0)^\top$ and $\epsilon_1 \defeq 1$.}
\label{fig:tilde_f}
\end{figure}

Note that our expression for $\tilde{f}$ is slightly simpler than the one in \cite{Malherbe2017}, while serving the same purpose, and it also corrects an issue in the original proof, which fails when $f$ is constant. A visual representation of this function is given in \cref{fig:tilde_f}.

Finally, let $\delta \defeq \max_{x \in \Omega} \tilde{f}(x) - \max_{x \in \Omega} f(x) > 0$. By construction of $\tilde{f}$, for any sequence of points $(X_i)_{0 \le i \le n}$ such that, for any $i \in \llbracket 0, n \rrbracket$, $X_i \notin B(c^\star, \epsilon_1 / 2)$ implies that
$$
\max_{x \in \Omega} \tilde{f}(x) - \max_{i = 0 \dots n} \tilde{f}(X_i) > \delta.
$$
Therefore, using \cref{eq:almost_never} and \cref{thm:eq_restrict}, we have that
$$
0 < \frac{\epsilon_2}{2 N_1} \le \mu_n^{(\tilde{f})} \left( \left\{ (X_i)_{0 \le i \le n} \; \middle| \; \max_{x \in \Omega} \tilde{f}(x) - \max_{i = 0 \dots n} \tilde{f}(X_i) > \delta \right\} \right),
$$
and thus
$$
  \mu_n^{(\tilde{f})} \left( \left\{ (X_i)_{0 \le i \le n} \middle| \max_{x \in \Omega} \tilde{f}(x) - \max_{i = 0 \dots n} \tilde{f}(X_i) > \delta \right\} \right) \xnrightarrow[n \to \infty]{} 0,
$$
which contradicts the fact that the algorithm is consistent on $\tilde{f}$.

The central role of our framework is evident in this proof, as \cref{thm:eq_restrict} and \cref{thm:fin_measure_mono} are indispensable tools for the formal verification of this result.

\section{Conclusion}
This work presents an abstract, measure‑theoretic framework that serves as a practical tool for reasoning about stochastic iterative global optimization algorithms. Built solely from an initial law and a sequence of probability kernels, the framework leverages the Ionescu–Tulcea theorem to produce probability measures on finite and infinite iteration sequences and yields useful and intuitive properties. This framework answers the need for a general, implementation‑independent formalism in the analysis of such algorithms, providing the foundation for formalizing general results in proof assistants.

To demonstrate its practical applicability, we showed that common algorithms (e.g., \textsc{adalipo}, \textsc{cma-es}, SDE‑based methods) fit naturally within the framework, and we used it to establish a clear, implementation‑independent proof of a general consistency theorem (Proposition~3 of \cite{Malherbe2017}). Critically, the entire framework and the proof of the proposition were formalized in the \texttt{Lean} proof assistant, ensuring the correctness of the definitions and proofs. This formalization effort demonstrates that rigorous verification of optimization theory is achievable and establishes a reusable foundation for future developments in the field.

By providing machine-verified building blocks, this work enables the inherent modularity of proof assistants in the optimization domain: definitions and theorems formalized here can be directly leveraged in subsequent formalizations, transforming algorithmic verification from isolated, ad hoc efforts into cumulative contributions that strengthen the collective infrastructure. Future work includes formalizing additional algorithms as concrete instances of the framework, establishing convergence rates and other algorithmic properties, and extending the framework to handle stochastic optimization in more general settings, such as unbounded domains or non-compact search spaces.

\section*{Acknowledgment}
We thank Étienne Marion for his assistance with the formalization of \cref{thm:eq_restrict} and for his work on the formalization of the Ionescu–Tulcea theorem in Mathlib, as well as all members of the \texttt{Lean} Zulip chat\footurl{https://leanprover.zulipchat.com/} for their support. The authors acknowledge the support from the Industrial Data Analytics and Machine Learning Chair hosted at ENS Paris-Saclay.

\bibliographystyle{plainnat_cap}
\bibliography{refs}%

\end{document}